%% file: prc.tex
\documentclass[aps,prc,twocolumn,showpacs,floatfix,nofootinbib,preprintnumbers,superscriptaddress,amsmath,amssymb,nofootinbib,longbibliography]{revtex4-1}

\usepackage{graphicx}
\usepackage{epsfig}
\usepackage{bm}
\usepackage{color}
\usepackage{float}
\usepackage{dcolumn}
\usepackage{multirow} 
\usepackage{hyperref}
\usepackage{dcolumn}

\graphicspath{{.}{figs/}}

\newcommand{\beq}{\begin{equation}}
\newcommand{\eeq}{\end{equation}}
\newcommand{\beqa}{\begin{eqnarray}}
\newcommand{\eeqa}{\end{eqnarray}}

\newcommand{\dbar}[1]{\overline{\overline{\mathcal{#1}}}}
\newcommand{\sbar}[1]{\overline{\mathcal{#1}}}

\begin{document}

\title{Shell-model coupled-cluster method for open-shell nuclei}

\author{Z.~H.~Sun} \affiliation{Department
  of Physics and Astronomy, University of Tennessee, Knoxville, TN
  37996, USA} \affiliation{Physics Division, Oak Ridge National
  Laboratory, Oak Ridge, TN 37831, USA} 

\author{T.~D.~Morris} \affiliation{Department
  of Physics and Astronomy, University of Tennessee, Knoxville, TN
  37996, USA} \affiliation{Physics Division, Oak Ridge National
  Laboratory, Oak Ridge, TN 37831, USA} 

\author{G.~Hagen} \affiliation{Physics Division, Oak Ridge National
  Laboratory, Oak Ridge, TN 37831, USA} \affiliation{Department of
  Physics and Astronomy, University of Tennessee, Knoxville, TN 37996,
  USA}

\author{G.~R.~Jansen} \affiliation{National Center for Computational Sciences, 
  Oak Ridge National Laboratory, Oak Ridge, TN 37831, USA}
  \affiliation{Physics Division, Oak Ridge National Laboratory, Oak Ridge, TN 37831, USA} 

\author{T.~Papenbrock} \affiliation{Department
  of Physics and Astronomy, University of Tennessee, Knoxville, TN
  37996, USA} \affiliation{Physics Division, Oak Ridge National
  Laboratory, Oak Ridge, TN 37831, USA}

\begin{abstract} 
 We present an approach to derive effective shell-model interactions
 from microscopic nuclear forces. The similarity-transformed
 coupled-cluster Hamiltonian decouples the single-reference state of a
 closed-shell nucleus and provides us with a core for the shell
 model. We use a second similarity transformation to decouple a
 shell-model space from the excluded space. We show that the
 three-body terms induced by both similarity transformations are
 crucial for an accurate computation of ground and excited
 states. As a proof of principle we use a nucleon-nucleon interaction
 from chiral effective field theory, employ a $^4$He core, and compute
 low-lying states of $^{6-8}$He and $^{6-8}$Li in $p$-shell model
 spaces. Our results agree with benchmarks from full configuration
 interaction.
\end{abstract}


\maketitle
\input{introduction}

\input{method}
\input{results}

\input{summary}

\begin{acknowledgments}
  This material is based upon work supported by the U.S. Department of
  Energy, Office of Science, Office of Nuclear Physics under Award
  Numbers DEFG02-96ER40963 (University of Tennessee), DE-SC0008499
  (SciDAC-3 NUCLEI), DE-SC0018223 (SciDAC-4 NUCLEI), DE-SC0015376
  (Double-Beta Decay Topical Collaboration), and the Field Work
  Proposals ERKBP57 and ERKBP72 at Oak Ridge National Laboratory
  (ORNL).  Computer time was provided by the Innovative and Novel
  Computational Impact on Theory and Experiment (INCITE) program. This
  research used resources of the Oak Ridge Leadership Computing
  Facility located at ORNL, which is supported by the Office of
  Science of the Department of Energy under Contract No.
  DE-AC05-00OR22725.
\end{acknowledgments}

%

\end{document}

%% file: introduction.tex
\section{Introduction}
Remarkable progress has been made in the ongoing endeavor to
understand and predict nuclei from underlying nuclear forces.  The
computations of atomic nuclei, starting from Hamiltonians with
nucleon-nucleon and three-nucleon forces are now starting to cover
significant portions of the nuclear landscape, from
light~\cite{pieper2001,navratil2009,barrett2013} to
medium-mass~\cite{roth2012,hagen2012b,soma2013b,lahde2014,hergert2014,stroberg2016,simonis2017,leistenschneider2018},
and to heavy nuclei~\cite{binder2013b,morris2018}. Theory now makes
predictions for processes and observables that are hard to
measure~\cite{quaglioni2008,epelbaum2011,elhatisari2015,hagen2015},
and for nuclei that are hard to produce~\cite{hagen2016b,morris2018}.
This progress is due to realistic nuclear
interactions~\cite{hebeler2011,roth2011a,ekstrom2015a} rooted in
chiral effective field theory~\cite{epelbaum2009,machleidt2011}, ideas
from effective field theories and the renormalization
group~\cite{bogner2003,bogner2010} and application of methods which
avoid the catastrophic scaling of full diagonalizations at the
acceptable cost of approximations
~\cite{dickhoff2004,gandolfi2007,lee2009,tsukiyama2011,binder2013,soma2013,hagen2014,hergert2016}.
Here, we mention in particular the coupled-cluster
(CC)~\cite{kuemmel1978,bartlett2007,hagen2014} and the in-medium
similarity renormalization group
(IMSRG)~\cite{tsukiyama2011,hergert2016} methods, which perform a
similarity transformation of the input Hamiltonian such that a product
state is decoupled from particle-hole excitations (i.e., it becomes an
exact eigenstate of the similarity transformed Hamiltonian).

In spite of all this progress, the description of open-shell nuclei
still poses challenges. For decades the nuclear shell
model~\cite{brown1988,caurier2005,shimizu2012} has been
the method of choice to address open-shell nuclei accessible in one
or two major oscillator shells. However, up until very recently, it
has been a challenge to tie the applied effective shell-model
interactions to the underlying nucleon-nucleon and three-nucleon
interactions. The widely used microscopic derivation of effective
interactions is usually based on many-body perturbation
theory~\cite{hjorthjensen1995,CORAGGIO2009135}, where certain diagrams
are summed up to infinite order by using the folded-diagram
method~\cite{KUO197165,RevModPhys.39.771}. However, the strong
correlations of nuclear systems and the tensor property of the nuclear
interactions spoil the order-by-order convergence of the effective
interaction in this
approach~\cite{BARRETT1970145,PhysRevC.7.1776,schucan1973}.
Alternative approaches to effective interactions contain both input
from the underlying nucleon-nucleon interaction and phenomenological
adjustments, which are difficult to make for large shell-model
spaces~\cite{honma2002,brown2006}. The idea to derive the effective
interaction from ab-initio methods~\cite{lisetskiy2008,dikmen2015} has
been fruitful: In this approach, one computes a doubly-magic nucleus
with mass number $A_c$, and extracts the effective interaction from
ab-initio calculations of nuclei with mass numbers $A=A_c + 1$ and
$A=A_c + 2$ using Lee-Suzuki-Okamoto (LSO)
techniques~\cite{suzuki1980,suzuki1982,suzuki1992,suzuki1994}. This
idea was implemented using CC methods in
Refs.~\cite{jansen2014,jansen2016}, respectively, starting from
nucleon-nucleon and three-nucleon forces. However, some observables,
such as charge radii, are sensitive to details of the high-lying
excited states in the $A=A_c + 2$ system that are used in the
construction of the effective interaction using the LSO projection
technique.

The valence-space IMSRG
(VS-IMSRG)~\cite{bogner2014,stroberg2016,stroberg2017} is an
alternative that avoids this problem. In this approach, a
non-perturbative effective interaction is derived within the IMSRG
formalism~\cite{tsukiyama2011}, i.e., by decoupling a core and a
valence space via similarity transformations. This avoids the problem
of projecting high-lying states onto the valence space using the LSO
technique, and, furthermore, the similarity transformation can be
consistently applied to observables other than the energy, potentially
offering insights into the origin of effective charges and the role of
two-body currents. The similarity transformation induces many-body
terms, and, at present, the VS-IMSRG and the IMSRG methods keep up to
normal-ordered two-body operators. The effect of this truncation is
currently not well understood.

In this paper, we want to follow the VS-IMSRG idea and generate
effective interactions and operators via a decoupling of the core from
the valence space. However, we base this decoupling, named shell-model
coupled cluster (SMCC), on the CC approach. Like the VS-IMSRG, the
SMCC is an extension of the closed-shell CC theory.  After decoupling
a closed-shell core with the CC method, we implement a second
similarity transformation of the Hamiltonian, which decouples the
valence space and the excluded space.  In contrast to the VS-IMSRG, we
perform this decoupling in a non-Hermitian way. This simplifies the
working equations that implement the transformation.  In this
simplified set of working equations we also explore the effect of
three-body terms in the similarity transformation.  We find that when
the leading order three-body terms are included in the calculation of
open $p$-shell nuclei, we obtain a much better agreement with exact
results than what is achieved with the SMCC or VS-IMSRG methods
truncated at the two-body level.

%% file: method.tex
\section {Shell-model coupled-cluster method} 
\subsection {Single reference coupled-cluster}

We start from the intrinsic Hamiltonian
\begin{equation}
\label{insh}
H=\left(1-\frac1{A}\right)\sum_{i=1}^A\frac{p_i^2}{2m}+\left( \sum_{i<j=1}^Av_{ij}
-\frac{\overrightarrow{p_i}\cdot \overrightarrow{p_j}}{mA} \right) .
\end{equation}
Here $A$ is the number of nucleons and $v_{ij}$ is
nucleon-nucleon potential.  We note that
if we are targeting an open system, then $A$ may 
be larger than the $A_c$, the number of core-nucleons addressed in 
closed-shell CC theory.

We normal order the Hamiltonian $H$ with respect to the
Hartree-Fock state
\begin{equation}
  |\Phi_0\rangle\equiv \prod_{i=1}^{A_c}a_i^\dagger |0\rangle
\end{equation}
and obtain
\begin{equation}
\mathcal{H}=H-E_{\text{ref}}=\sum_{pq}f_{pq}\{p^\dagger q\}+\frac14\sum_{pqrs}V_{pqrs} \{p^\dagger q^\dagger sr\}.
\end{equation}  
Here, the brackets $\{\cdots \}$ indicate normal ordering,
$E_{\text{ref}}$ is the HF energy, $f$ is the one-body interaction
(also known as the Fock-matrix), $V$ is the anti-symmetric two-body
interaction, and $p,q,r,s$ label single-particle states, see, e.g.,
Ref.~\cite{shavittbartlett2009} for details.  In single-reference CC
theory~\cite{kuemmel1978,bartlett2007,hagen2014} , the ground-state
state is given by the exponential ansatz,
\begin{equation}
|\Psi_0\rangle=\exp{(T)}|\Phi_0\rangle.
\end{equation} 
Many-body correlations are introduced via the cluster operator
\begin{equation}
  T=T_1+T_2+T_3+\cdots+T_{A_c} .
  \label{Top}
\end{equation}
Here the cluster operators $T_n$ create $n$-particle--$n$-hole
excitations
 \begin{equation}
 T_n=\left(\frac1{n!}\right)^2\sum_{\substack{ a_1a_2\cdots a_n \\  i_1i_2\cdots i_n}} t^{a_1a_2\cdots a_n}_{i_1i_2\cdots i_n} \{a_1^\dagger a_2^\dagger\cdots a_n^\dagger i_n i_{n-1}\cdots i_1\},
\label{eq:tamps}
 \end{equation}
 and $t^{ a_1a_2\cdots a_n}_{i_1i_2\cdots i_n}$ are the corresponding
 amplitudes that need to be determined.  The CC method yields the
 similarity-transformed normal-ordered Hamiltonian
\begin{eqnarray}
\overline{\mathcal{H}}&\equiv&\exp{(-T)}\mathcal{H}\exp{(T)} \notag \\
&=&\left(\mathcal{H}\exp{(T)}\right)_C \notag \\
&=& \Delta E+\sum_{pq}\overline{\mathcal{H}}_{pq}\{p^\dagger q\}+\sum_{pqrs} \overline{\mathcal{H}}_{pqrs}\{p^\dagger q^\dagger sr\}\nonumber \notag \\
&\quad&+\sum_{pqrstu} \overline{\mathcal{H}}_{pqrstu}\{p^\dagger q^\dagger r^\dagger uts\}+\cdots . \label{hccsd}
\end{eqnarray}
Here, $\Delta E$ is the correlation energy, i.e., the correction to the
HF energy, and $\overline{\mathcal{H}}_{pq}$
,$\overline{\mathcal{H}}_{pqrs}$ and $\overline{\mathcal{H}}_{pqrstu}$
are the one-, two-, and three-body part of $\overline{\mathcal{H}}$,
respectively.  The subscript $C$ in the second line of
Eq.~(\ref{hccsd}) indicates that only connected diagrams enter the
expression. The CC correlation energy and the amplitudes in
Eq.~(\ref{eq:tamps}) are determined by solving the following
equations,
\begin{eqnarray}
\langle \Phi_0|\overline{\mathcal{H}}|\Phi_0\rangle&=&\Delta E , \\
\langle \Phi^a_i|\overline{\mathcal{H}}|\Phi_0\rangle&=&0  , \label{ccsingles}\\
\langle \Phi^{ab}_{ij}|\overline{\mathcal{H}}|\Phi_0\rangle&=&0 ,  \label{ccdoubles} \\
\langle
\Phi^{abc}_{ijk}|\overline{\mathcal{H}}|\Phi_0\rangle&=&0 \label{cctriples}. \\
\nonumber
\vdots
\end{eqnarray}
So far we have not introduced any approximations. However, in
practical calculations we need to truncate the cluster operator $T$ at
some low-order rank. The most commonly used approximation is CC with
singles and doubles excitations (CCSD). In the CCSD approximation only
Eqs.~(\ref{ccsingles})-(\ref{ccdoubles}) are solved to determine the
amplitudes. In this work we go beyond the CCSD approximation and
include the leading-order three-particle-three-hole excitations in
Eq.~(\ref{cctriples}). This approximation is termed
CCSDT-1~\cite{lee1984} and gives a significant improvement in the
description of the ground-state. For closed shell systems, CCSD and
CCSDT-1 typically capture around 90\% and 99\% of the full correlation
energy, respectively~\cite{bartlett2007}.

\subsection {Open-shell coupled-cluster}
We now turn to the derivation of an effective shell-model interaction
in our SMCC approach.  We identify a valence space outside a closed
shell core, and then divide the Hilbert space of
$\overline{\mathcal{H}}$ into to two complementary spaces $P$ and $Q$
with
 \begin{equation}
 P+Q=I.
\end{equation}  
Here the operator $P$ is a projector onto the core and valence space,
and $Q$ is the excluded space. We seek a second similarity
transformation that decouples the $P$ and $Q$ spaces, i.e.,
\begin{equation}
\label{hbb}
\overline{\overline{\mathcal{H}}}\equiv\exp{(-S)}\overline{\mathcal{H}}\exp{(S)},
\end{equation}
such that
\begin{equation}
\label{decp}
Q\overline{\overline{\mathcal{H}}} P=0
\end{equation}
with
\begin{equation}
S=QSP.
\end{equation}

In the second similarity transformation given in Eq.~(\ref{hbb}), we
only retain the one- and two-body parts of the initial CCSDT-1
similarity transformed Hamiltonian $\overline{\mathcal{H}}$. The
transformation fulfilling Eqs.~(\ref{hbb}) and (\ref{decp}) yields a
non-Hermitian effective interaction
\begin{equation}
\label{hveff}
H_{\text{eff}}=P\overline{\overline{\mathcal{H}}}P=P\exp{(-S)}\overline{\mathcal{H}}\exp{(S)}P.
\end{equation}
In analogy to the cluster operator $T$ of Eq.~(\ref{Top}), the
operator $S$ links the excluded space ($Q$) and the model space ($P$)
and in general contains up to $A$-body terms.

\begin{figure}[ht]
\includegraphics[width=0.4\textwidth]{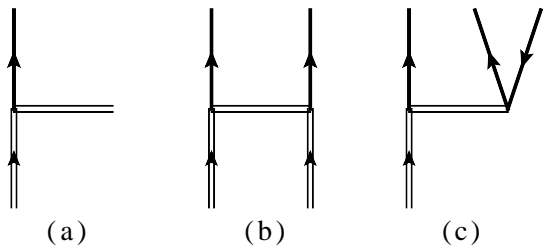}
\caption{Diagrams of the operator $S$ that couple the valence space
  and the excluded space. The horizontal double bar is the $S$
  operator, with incoming and outgoing particle lines as indicated by
  arrows. The double incoming line denotes a particle inside the model
  space. Diagram (a) is a one-body operator, while diagrams (b) and
  (c) are two-body operators. The solid particle line is a general
  particle (i.e., either from the excluded space or the model space),
  except for the outgoing two particles in diagram (b), where at least
  one of the outgoing particles must be from the excluded space.}
\label{fig:qp}
\end{figure}

Fig.~\ref{fig:qp} shows the three topologies of $S$ that are found
up to the two-body level.  We notice that if one starts from the bare
Hamiltonian in Eq.~(\ref{insh}), there would be additional terms with
a cluster structure.  These are not present in the core-decoupled CC
Hamiltonian $\overline{\mathcal{H}}$.  The diagrams in
Fig.~\ref{fig:qp} correspond to the amplitudes $S_{ab}$, $S_{abcd }$
and $S_{abci}$, where we used labels $a, b, c,\ldots $ for particle
states, and $i, j, k,\ldots$ for hole states.

To determine the amplitudes of $S$ proves to be less straightforward
than doing so for the $T$ amplitudes. Similar equations as those found
in Eqs.~(\ref{ccsingles})-(\ref{ccdoubles}) arise from
Eq.~(\ref{decp}),
\begin{eqnarray}
\overline{\overline{\mathcal{H}}}_{ab}&=0 \label{eq:Sab}\, , \\
\overline{\overline{\mathcal{H}}}_{abci}&=0 \label{eq:Sabic}\, ,\\ 
\overline{\overline{\mathcal{H}}}_{abcd}&=0 \label{eq:Sabcd}\, .
\end{eqnarray}
Inspired by the Magnus formulation of the IMSRG
method~\cite{morris2015}, we adopt a constructive approach of building
up $S$ and solving Eqs.~(\ref{eq:Sab})-(\ref{eq:Sabcd}).

In the original formulation of the IMSRG, a unitary transformation
which depends on a continuous variable $t$ is introduced, making a
"flowing" transformed Hamiltonian,
\begin{eqnarray}
H(t)=U^\dagger(t)H U(t) \, .
\end{eqnarray} 
The derivative of $H(t)$ with respect to $t$, denoted as $\dot{H}(t)$,
yields
\begin{eqnarray}
\dot{H}(t)= [\eta(t),H(t)] \label{eq:floweq}\\
\eta(t) = \dot{U}^\dagger(t)U(t)\label{eq:eta}.
\end{eqnarray} 
Here $\eta$ is chosen such that it eliminates the undesired matrix
elements, and does so in a way that does not give rise to exceedingly
stiff equations.  With a suitable choice of $\eta$ determined,
Eq.~(\ref{eq:floweq}) is solved until the undesired matrix elements
are suppressed to an acceptable criteria.  The Magnus formulation of
the IMSRG dispenses with solving Eq.~(\ref{eq:floweq}), and
instead exploits the fact that Eq.~(\ref{eq:eta}) allows for a direct
solution of the transformation $U(t)$ via the following expressions
\begin{eqnarray}
U(t)^\dagger &=& \exp(\Omega(t))  \label{eq:expomega} \, ,\\
\dot{\Omega}(t) &=& \eta(t) - \frac{1}{2}[\Omega(t),\eta(t)]+ \dots \label{eq:domega}\, .
\end{eqnarray}
There are several benefits to solving Eq.~(\ref{eq:domega}), most
notably that observables can be solved one-by-one after a converged
$\Omega$ is obtained.  This is opposed to having to solve every
desired observable other than the energy in parallel to the
Hamiltonian, which would be prohibitively expensive for almost any
single tensor operator except for a few scalar operators such as
radii.

Taking inspiration from this Magnus formulation of the IMSRG, we solve
an equation similar to Eq.~(\ref{eq:domega}) for our operator $S$,
\begin{equation}\label{sode}
\dot{S}(t)=-\eta\left(\dbar{H}(t)\right).
\end{equation}
We have omitted all higher order terms in Eq.~(\ref{eq:domega})
because we only want to keep the form of $S$ as it originates from
Eqs.~(\ref{eq:Sab})-(\ref{eq:Sabcd}), and it is only important to find
$S$ that decouples the $P$ and $Q$ space, not what path is taken by
the solver to arrive at the solution.  We use the same definition of
$\eta$ as is used in the IMSRG except that our $\eta$ is not
anti-Hermitian.  Its form will be discussed below.  As $\eta$ is a
function of the Hamiltonian, one must generate the transformed
Hamiltonian at each time step in order to decide which direction to
take $S$.  This is accomplished via the infinite
Baker-Campbell-Hausdorff (BCH) expansion, and we truncate the expansion
when nested commutators have matrix norms below a small threshold.  Thus,
\begin{eqnarray}\label{magnus}
\dbar{H}&=&\sbar{H}+\left[\sbar{H},S\right]+\frac1{2!}\left[\left[\sbar{H},S\right],S \right]+\cdots \nonumber \\
&=& \sum_{k=0}^\infty \frac1{k!}{\rm ad}^k_S(\sbar{H}).
\end{eqnarray}
Here ${\rm ad}^0_S(\sbar{H})=\sbar{H}$ and ${\rm
  ad}^k_S(\sbar{H})=\left[{\rm ad}^{k-1}_S(\sbar{H}),S\right]$. Starting from
two- and three-body Hamiltonians, the nested commutators generate
operators of increasing rank, and we need to introduce
truncations. These will be discussed below.

There are several choices for the generator in Eq.~(\ref{sode})
similar to the IMSRG~\cite{hergert2016}, and in the current work we
used the ``White'' generator~\cite{white2002}
\begin{equation}
\eta_{pqrs}=-\frac{\dbar{H}_{pqrs}}{\dbar{H}_{pp}+\dbar{H}_{qq}-\dbar{H}_{rr}-\dbar{H}_{ss}},
\end{equation}
and the ``Arc-tan White'' generator
\begin{equation}
\eta_{pqrs}=-\frac12\arctan\left( \frac{2\dbar{H}_{pqrs}}{\dbar{H}_{pp}+\dbar{H}_{qq}-\dbar{H}_{rr}-\dbar{H}_{ss}}\right).
\end{equation}

\begin{figure}[ht]
\includegraphics[width=0.45\textwidth]{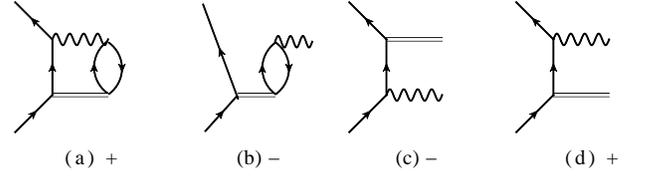}
\caption{Diagrammatic representation of $[\chi,S]^{\text{1b}}$
  multiplied with a $+$ or $-$ sign from the commutator as indicated
  below each diagram. The wiggly line refers to the intermediate
  $\chi$, and the horizontal double bar is $S$.}
\label{fig:s1}
\end{figure}

\begin{figure}[ht]
\includegraphics[width=0.45\textwidth]{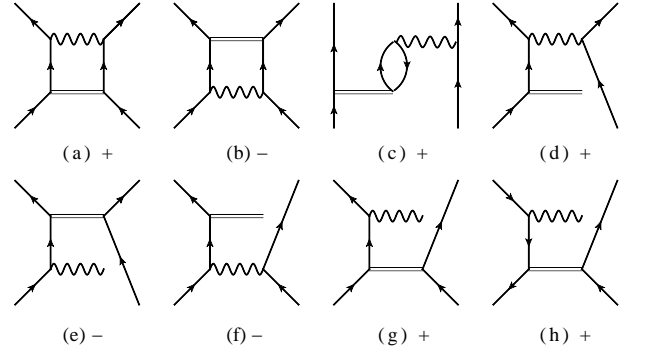}
\caption{Diagrammatic representation of $[\chi,S]^{\text{2b}}$ multiplied
  with a $+$ or $-$ sign from the commutator.}
\label{fig:s2}
\end{figure}

We first consider a truncation of Eq.~(\ref{magnus}) at the normal
ordered two-body (NO2b) level, and refer to it as SMCC(2b). As we use
nested commutators to evaluate the transformation leading to
$\dbar{H}$, we need only one general commutator expression for
non-trivial contributions from two non-Hermitian operators $\chi$ and
$S$.  The one- and two-body contributions from the commutator are
shown in Figs.~\ref{fig:s1} and \ref{fig:s2}, respectively. They lead
to the following algebraic expressions for the one-body terms
\begin{eqnarray}
[\chi,S]_{pq}&=& \frac12\sum_{abi}\chi_{piba}S_{baqi}\nonumber\\
                     &\quad &+\sum_{ia}\chi_{ia}S_{paqi}\nonumber\\
                      &\quad & +\sum_a(\chi_{pa}S_{aq}-S_{pa}\chi_{aq}),
\end{eqnarray}
and the two-body terms
\begin{eqnarray}
[\chi,S]_{pqrs}&=&-\frac12\sum_{ab}\left(\chi_{abrs}S_{pqab}-S_{abrs}\chi_{pqab}\right)\nonumber\\
                        &\quad &-(1-P_{pq})(1-P_{rs})\sum_{ia}\chi_{pias}S_{aqri}\nonumber\\
                        &\quad &+(1-P_{pq})\sum_a (\chi_{pa}S_{aqrs}-S_{pa}\chi_{aqrs})\nonumber\\
                        &\quad &-(1-P_{rs})\sum_a (\chi_{ar}S_{pqas}-S_{ar}\chi_{pqas})\nonumber\\
                        &\quad &-(1-P_{rs})\sum_i\chi_{ir}S_{pqis}.
\end{eqnarray}
The matrix elements are antisymmetric through exchange of
two equivalent external lines.  

\begin{figure}[ht]
\includegraphics[width=0.45\textwidth]{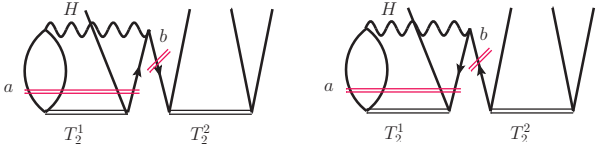}
\caption{(Color online) Two-body diagrams $[[\mathcal{H},T_2],T_2]$ in CCSD
  closed-shell calculation. (Here, the black double horizontal bar denotes
  the cluster operator $T$, while the wiggly line is the Hamiltonian $\mathcal{H}$.)
  The intermediate 3-body $[[\mathcal{H},T_2]_{3b},T_2]_{2b}$ and one-body
  $[[\mathcal{H},T_2]_{1b},T_2]_{2b}$ terms inside the commutator result from
  making cuts at the locations a and b, respectively, and are
  indicated by red double lines.}
\label{fig:ht2}
\end{figure}

While truncating all intermediates at the NO2b level is expedient, the
truncation comes at the cost of neglecting higher-body terms. It is
instructive is to look at the ground-state CC decoupling performed in
this approximation.  The similarity-transformed Hamiltonian can be
written in two ways (as expressed in the following two lines) 
\begin{eqnarray}
\overline{\mathcal{H}} &=& \left[\mathcal{H}\left(1+T+\frac{1}{2!}T^2+\frac{1}{3!}T^3+\frac{1}{4!}T^4\right)\right]_C\label{eq:hccsdexplicit1} \\
&=& \mathcal{H}+[\mathcal{H},T]+\frac{1}{2!}[[\mathcal{H},T],T]+\ldots \label{eq:hccsdexplicit2} .
\end{eqnarray}
If we apply the NO2b approximation to each commutator in
Eq.~(\ref{eq:hccsdexplicit2}), we find some large omitted terms
compared to Eq.~(\ref{eq:hccsdexplicit1}). For example, a term
contributing to $\frac12\mathcal{H}T_2^2$ arises in equal parts from
$[[\mathcal{H},T_2]_{3b},T_2]_{2b}$ and from
$[[\mathcal{H},T_2]_{1b},T_2]_{2b}$.  Here, the subscript $2b$ and
$3b$ on the commutator denotes the truncation at the two- and
three-body level, respectively, see Fig.~\ref{fig:ht2}. Thus using the
NO2b truncation, one undercounts this class of terms by a factor of
$\frac{1}{2}$. The IMSRG(2) approximation for ground-states also
neglects such terms. These terms are repulsive, while perturbative
triples are attractive.  Thus, this undercounting explains
why IMSRG(2) ground-state energies generally fall closer to the
CCSD(T) approximation than to CCSD~\cite{hergert2016}.  For this
example it is clear that the effect of
$[[\mathcal{H},T_2]_{3b},T_2]_{2b}$ can be restored simply in an
ad-hoc fashion by double counting the half of these terms that arises
from $[[\mathcal{H},T_2]_{1b},T_2]_{2b}$.

\begin{figure}[ht]
\includegraphics[width=0.4\textwidth]{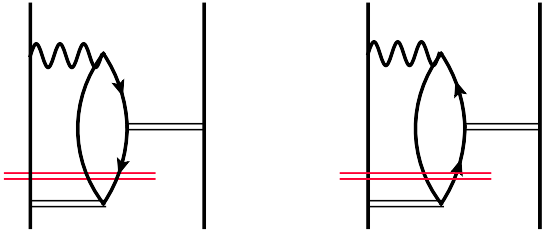}
\caption{(Color online) ``Core polarization'' diagrams that are
  neglected when all commutators are truncated at the two-body
  level. Cutting the diagrams at the red double bars reveals
  intermediate three-body terms.}
\label{fig:cor}
\end{figure}

A similar situation appears in the open-shell
decoupling. Fig,~\ref{fig:cor} shows a pair of ``core polarization''
diagrams in the open shell calculation that are neglected if the
commutators are kept only at the two-body level. Cutting the diagrams
as indicated at the red horizontal double bars clearly yields
three-body terms. Numerical evaluations demonstrate that these
diagrams, like their ground-state counterparts, make a substantial
difference in the valence-space decoupling.

\begin{figure}[ht]
\includegraphics[width=0.45\textwidth]{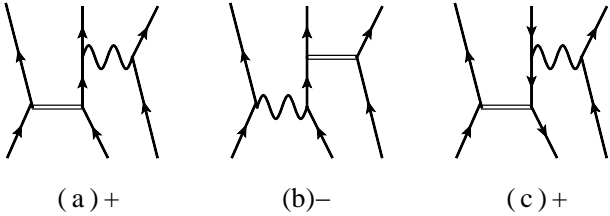}
\caption{Diagrammatic representation of $[\chi,S]_{\text{3b}}$
  multiplied with a $+$ or $-$ sign from the commutator as
  indicated. Here $\chi$ and $S$ are shown as wiggly and double
  horizontal lines, respectively.}
\label{fig:h3}
\end{figure}

The inclusion of commutators with intermediate three-body terms is
challenging because they should not affect the scaling of the storage
required for our calculation. We proceed as follows.  In each
commutator evaluation, the generated three-body diagrams are shown in
Fig.~\ref{fig:h3}.

\begin{figure}[ht]
\includegraphics[width=0.45\textwidth]{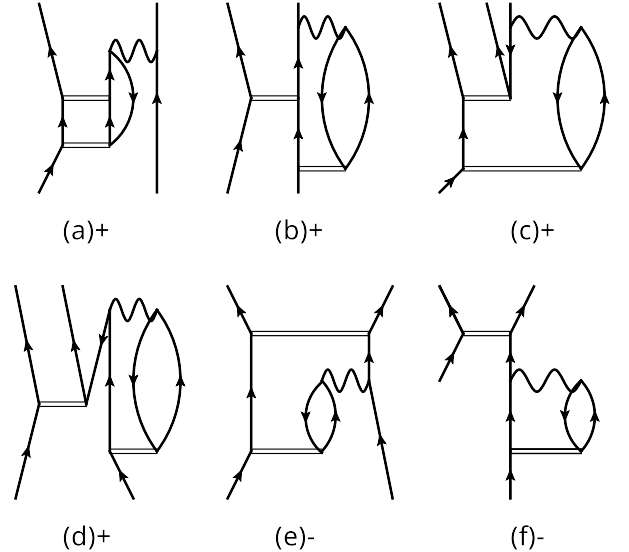}
\caption{Diagrams of $[[\chi_{2b},S_{2b}]_{3b},S_{2b}]_{2b}$
  multiplied with a $+$ or $-$ sign from the commutator as indicated below each diagram. Here $\chi$ and $S$ are shown as wiggly and double
  horizontal lines, respectively.}
\label{fig:h3s}
\end{figure}

They have the algebraic form 
\begin{eqnarray}
\chi_{pqrstu}&=&{\cal A}\sum_a\left( \overline{\mathcal{H}}_{qrau}S_{past}- S_{qrau}\overline{\mathcal{H}}_{past}\right) \nonumber \\
                  &\quad &+\sum_i\overline{\mathcal{H}}_{ipts}S_{rqui},
\end{eqnarray}
and the anti-symmetrizer is
\begin{equation}
{\cal A}\equiv(1-P_{pq}-P_{pr})(1-P_{us}-P_{ut}) .
\end{equation}
While seeking the solution of Eq.~(\ref{decp}), we use the above
expressions to keep all intermediate terms of the form
$[[\sbar{H}_{2b},S_{2b}]_{3b},S_{2b}]_{2b}$.  These terms are
calculated on the fly without changing the scaling, since all of them
can be factored into effective two-body pieces.
Fig.~\ref{fig:h3s} shows the diagrams corresponding to these
three-body terms we need to keep.  Their algebraic form are found in
Eq.~(\ref{eq:2b2b3b2b}). This approximation is labeled by
SMCC(3b-diag), as it effectively keeps diagonal contributions from the
three-body terms.
 \begin{eqnarray}\label{eq:2b2b3b2b}
[[\chi,S]_{3b},S]_{pqrs}&=&\frac12(1-P_{pq})(1-P_{rs})\sum_{abci}S_{bari}S_{pcba}\chi_{iqcs}\nonumber\\
 &\quad & +(1-P_{pq})(1-P_{rs})\sum_{abci}S_{basi}S_{pcrb}\chi_{qica}\nonumber\\
 &\quad & -(1-P_{rs})\sum_{abij}S_{bari}S_{pqbj}\chi_{ijas}\nonumber\\
  &\quad & -\frac12 (1-P_{rs})\sum_{abij}S_{basi}S_{pqrj}\chi_{jiba}\nonumber\\
   &\quad & -(1-P_{rs})\sum_{abci}S_{abir}\chi_{icas}S_{pqbc}\nonumber\\
    &\quad & -\frac12 (1-P_{rs})\sum_{abic}S_{basi}\chi_{ciba}S_{pqrc}\nonumber.
\end{eqnarray} 
Contributions from
$[[\sbar{H}_{2b},S_{2b}]_{3b},S_{2b}]_{3b},S_{2b}]_{2b}$ are neglected
  as they are too expensive to solve at each time step, and assumed to
  be of higher order.

Our tests below show that shell-model interactions produced at the
SMCC(3b-diag) approximation are still not satisfactory when compared
to exact diagonalizations. We recall that an improved accuracy for the
ground-state decoupling in CC theory requires us to include the
leading order $T_3$ that is linear in $T_2$. In the same spirit, we
can generate the leading order part of $S_3$ by performing a third
similarity transformation, where we keep only terms linear in $S_2$
that contribute to $S_3$.  The inclusion of an approximation for $S_3$
is necessary, because there are three-body pieces of
$\overline{\overline{\mathcal{H}}}$ that violate the decoupling
condition~(\ref{decp}). If we keep only those parts linear in
$S_{2b}$, then the elements of $S_{3b}$ can be found as the
approximate solutions of the following equations,
\begin{eqnarray}
Q \exp(-S_{3b})\overline{\overline{\mathcal{H}}}\exp(S_{3b})P &\approx& 0 \label{eq:S3}\, ,\\
\left\{\left[\overline{\mathcal{H}}_{2b},S_{2b}\right]+\left[\overline{\overline{\mathcal{H}}}_{1b},S_{3b}\right] \right\}_{abcdij}&=&0 \label{eq:Sabcdij}\, , \\
\left\{\left[\overline{\mathcal{H}}_{2b},S_{2b}\right]+\left[\overline{\overline{\mathcal{H}}}_{1b},S_{3b}\right] \right\}_{abcdei}&=&0 \label{eq:Sabcdei}\, ,\\ 
\left\{\left[\overline{\mathcal{H}}_{2b},S_{2b}\right]+\left[\overline{\overline{\mathcal{H}}}_{1b},S_{3b}\right] \right\}_{abcdef}&=&0 \label{eq:Sabcdef}\, .
\end{eqnarray}
If we further approximate $\overline{\overline{\mathcal{H}}}_{1b}$
above as containing only the diagonal one-body energies, then each of
the three three-body topologies can be solved for generally as
\begin{equation}
  \label{Spqrstu}
S_{pqrstu}=-\frac{[\overline{\mathcal{H}}_{2b},S_{2b}]_{pqrstu}}{\overline{\overline{\mathcal{H}}}_{pp}+\overline{\overline{\mathcal{H}}}_{qq}+
\overline{\overline{\mathcal{H}}}_{rr}-\overline{\overline{\mathcal{H}}}_{ss}
-\overline{\overline{\mathcal{H}}}_{tt}-\overline{\overline{\mathcal{H}}}_{uu}}.
\end{equation}

\begin{figure}[ht]
\includegraphics[width=0.45\textwidth]{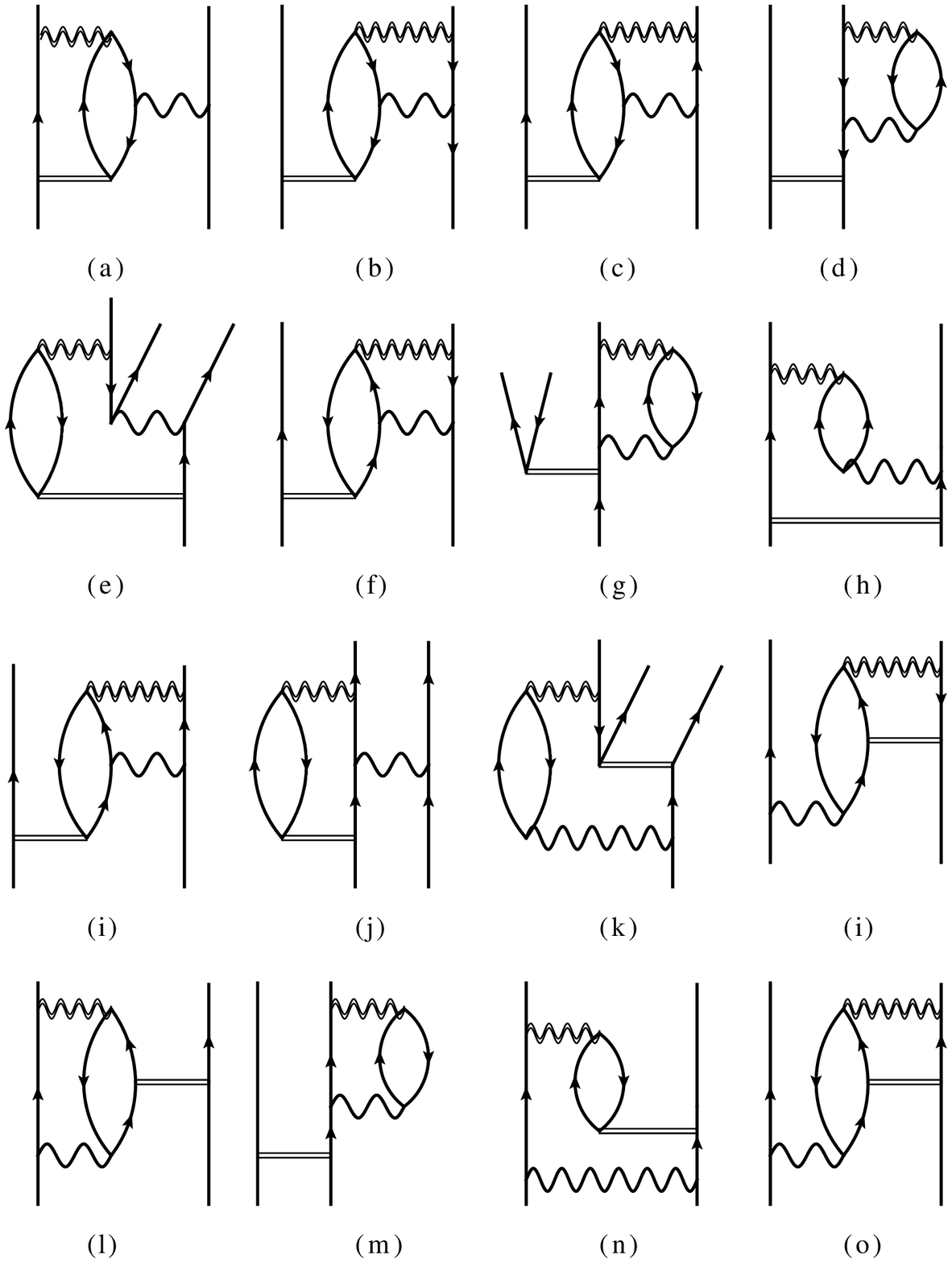}
\caption{Diagram representation of $[\dbar{H}_{2b},S_{3b}]_{2b}$.  }
\label{fig:s3}
\end{figure}

\begin{figure}[ht]
\includegraphics[width=0.45\textwidth]{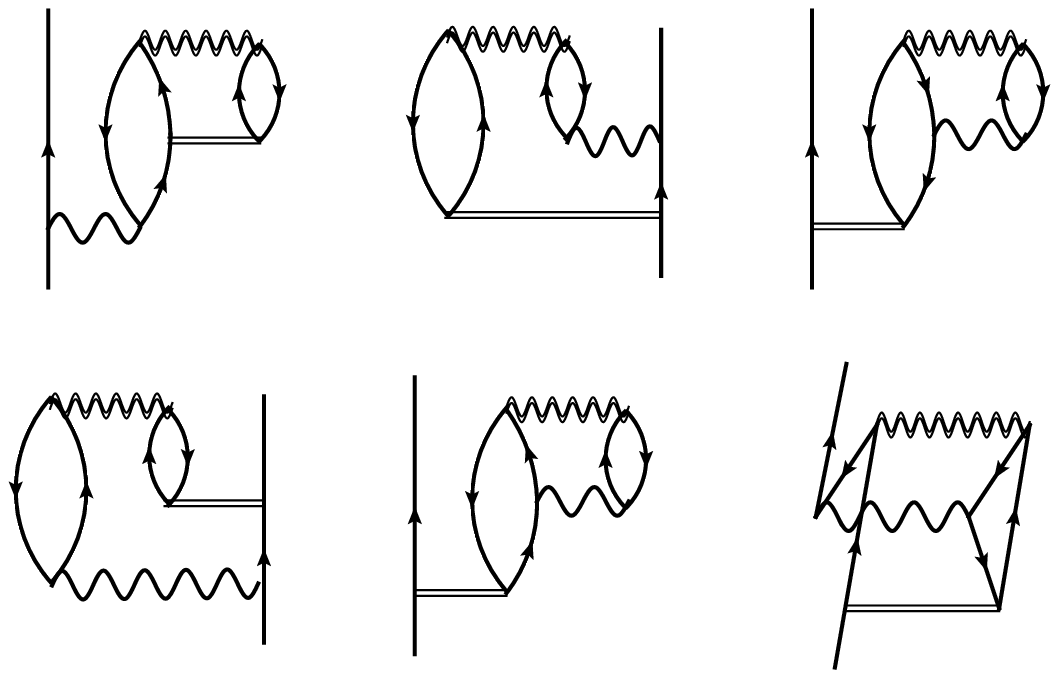}
\caption{Diagram representation of $[\dbar{H}_{2b},S_{3b}]_{1b}$. }
\label{fig:s31}
\end{figure}

Our goal is to generate an effective shell-model interaction, and we
wish to study how contributions from $S_3$ will affect this
interaction. For this purpose we generate the one- and two-body
contributions of
$\Delta\overline{\overline{\mathcal{H}}}=P[\overline{\overline{\mathcal{H}}},S_{3b}]P$,
and add them to the final valence interaction, and we refer to this as
SMCC(3b-od), because it keeps ``three-body off-diagonal'' terms.  We
perform this operation once after converged $S_{1b}$,$S_{2b}$ are
obtained. This is akin to a perturbative treatment. If we were instead
to solve the above equations iteratively, it would be akin to the
CCSDT-1 process for decoupling the ground-state in CC theory. The
elements we keep from $\Delta\overline{\overline{\mathcal{H}}}$ can be
organized into $\left[\overline{\overline{\mathcal{H}}}_{2b},S_{3b}\right]_{2b}$
and $\left[\overline{\overline{\mathcal{H}}}_{2b},S_{3b}\right]_{1b}$, and the
corresponding diagrams are shown in Figs.~\ref{fig:s3} and
\ref{fig:s31}, respectively. In each diagram, the short-wavelength
small-amplitude wiggly line represents $\dbar{H}$, while the remaining
diagram represents the three-body term~(\ref{Spqrstu}) that is made
from combining the three-body contraction $\left[\overline{\mathcal
    H}_{2b},S_{2b}\right]_{3b}$ with the energy denominator.


In addition to the perturbative treatment of $S_{3b}$, we also keep
the leading order three-body valence interaction of
$\overline{\overline{H}}_{3b}\approx\left[\overline{H},S_{2b}\right]_{3b}$. This
is important as the three-body valence interaction that we include
would appear at second-order in the interaction.  This procedure leads
to a one-, two-, and three-body valence-space interaction, and we
refer to this as the SMCC(3b) approximation.

\subsection{Expectation values of observables in SMCC}

As the similarity transformed Hamiltonian is not Hermitian when
computed with the CC method, the evaluation of ground-state
expectation values of operators other than the energy requires one to
compute the left ground-state. This approach is based on response functions~\cite{salter1989}.
The left ground-state eigenvector is determined by
\begin{equation}\label{lvec}
\langle \Phi_0 |(1+\Lambda)(\overline{\mathcal{H}}-\Delta E)=0,
\end{equation}
where $\Lambda$ is a de-excitation operator.
The expectation value of a ground-state observable is then
\begin{equation}
O_{c} = \langle \Phi_0|(1+\Lambda)\overline{\mathcal{O}}|\Phi_0\rangle.
\end{equation}
For our valence-space approach, right states have the exact form of
\begin{equation}
|R_\nu\rangle=|\Phi_0\rangle\otimes|\Psi_\nu\rangle .
\end{equation}
Here $|\Psi_\nu\rangle$ is the wave function obtained from a shell-model diagonalization. 
For the left, the interplay between the core and valence space are neglected. However,
the terms affecting observables should still be adequately captured by approximating
the left eigenvector as
\begin{equation}
\langle L_\nu|=\langle\Phi_0|(1+\Lambda)|\otimes\langle\tilde{\Psi}_v| ,  
\end{equation}
where $\langle\tilde{\Psi}_v|$ is a wave function obtained from an left 
diagonalization of $P\dbar{H}P$.
This approximate expression makes it possible to deal with the core
and valence space separately, and the expectation value of observables
are calculated as
\begin{equation}
\langle L_\nu |O| R_\nu\rangle= O_c +\langle \tilde{\Psi}_v|O_v |\Psi_v\rangle .
\end{equation}
In what follows, we will use this expression to compute ground-state
radii.

%% file: results.tex
\section{Results and benchmarks}
We present results for the $p$-shell nuclei $^{6,7,8}$He and
$^{6,7,8}$Li from the SMCC(2b) and SMCC(3b), and compare them with
exact results from the full configuration interaction (FCI) approach,
and with other approaches based on CC theory and IMSRG.  We use the
chiral nucleon-nucleon N$^3$LO interaction from \textcite{entem2003}.  The
bare interaction employs a cutoff at $\Lambda=500$~MeV and is softened
via the free-space similarity renormalization group method to the
cutoff 1.9~fm$^{-1}$~\cite{bogner2007}. We neglect any three-body
forces in this initial study in order to focus on the effects of the
many-body terms induced by the valence-space decoupling. The two-body
matrix elements are calculated in the spherical harmonic oscillator
basis with $\hbar\omega=24$~MeV. The model space consist of five major
oscillator shells with single-particle angular momenta up to and
including $l_{\text{max}}=2$. This space consists of 76 single
particle states and allows us to perform proof-of-principle and
benchmark calculations.

The FCI results are obtained from an exact diagonalization of the
Hamiltonian matrix. The coupled-cluster approach to the $A=6,7,8$
nuclei is as follows. The $A=6$ nucleus $^6$He is computed with the
two-particle attached equation-of-motion (EOM) coupled-cluster method
by attaching two neutrons to the $^4$He core. That approach includes
$2p$-$0h$ and $3p$-$1h$
configurations~\cite{jansen2011,jansen2012,gour2008}. The $^8$He
nucleus exhibits a closed sub shell and can be computed directly with
coupled-cluster theory. Its $^8$Li isobar is obtained from a
charge-exchange EOM approach~\cite{ekstrom2014}. The VS-IMSRG
computations of the $A=6,7,8$ proceed as presented in
\cite{stroberg2017}: The $^4$He core is decoupled in the IMSRG(2)
approximation and then followed by a valence-space decoupling and
valence-space diagonalization using NushellX~\cite{brown2014}.

\begingroup
\squeezetable
\begin{table}[]
\centering
\begin{tabular}{cccccc}
\hline
\hline
       &  IMSRG(2)&   CCSD(2b) &   CCSD(3b-diag)&   CCSD   &   CCSDT-1  \\
E(MeV) &  -24.573 &   -24.626&  -23.912&   -23.911&   -24.261   \\
\hline
\hline                
\end{tabular}
\caption{Ground-state energy of the $^4$He core in a $^6$He from CCSD variants
  and IMSRG(2). \label{table1}}
\end{table}
\endgroup
Let us first study the effect of induced three-body terms in the
ground-state decoupling of the $^4$He core.  We calculated the energy
of the $^4$He core used in a $^6$He diagonalization, employing
IMSRG(2), CCSD(2b) [where each nested commutator keeps only two-body
  terms in analogy to SMCC(2b)], CCSD(3b-diag) [where diagonal
  three-body terms are kept in the spirit of SMCC(3b-diag)], CCSD, and
CCSDT-1. Table~\ref{table1} shows the results.  CCSD(2b) is about 700
keV overbound, and qualitatively similar to the IMSRG(2) result
supporting our claim that it is an undercounting of terms in
Fig.~\ref{fig:ht2} that separates IMSRG(2) from CCSD.  We see that
keeping $[[H,T_2]_{3b},T]_{2b}$, in the CCSD(3b-diag) approximation
almost matches full CCSD.  This also suggests that our evaluation of
the BCH expansion using nested commutators is appropriate.
Calculations performed at the SMCC(3b) level yield similar results to
CCSDT-1.

\begin{figure}[ht]
\includegraphics[width=0.5\textwidth]{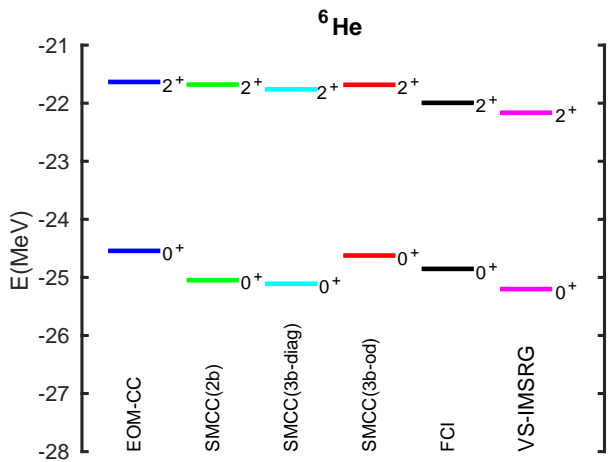}
\caption{(Color online) Spectrum of $^6$He computed with various
  methods (FCI, EOM-CC and IMSRG) and compared to the SMCC
  results.}
\label{fig:he6}
\end{figure}

For the open-shell calculation of $^6$He, we start from the one- and
two-body CCSDT-1 Hamiltonian. This Hamiltonian has the $^4$He core
decoupled at the CCSDT-1 approximation; the induced three-body terms
from CC are neglected.  The $p$-shell effective interaction is
obtained by decoupling the Hamiltonian with the SMCC method in the
approximations discussed in the previous Section.  The resulting
shell-model interaction is used to calculate $^6$He, and we compare
those results with FCI. In the $p$-shell, $^6$He is a two
valence-neutron system with configurations
$|(0p\frac12)^2\rangle^{0^+}$, $|(0p\frac32)^2\rangle^{0^+,2^+}$ and
$|(0p\frac12 0p\frac32)\rangle^{1^+,2^+}$. The occupations of the FCI
wave functions show that the $0_1^+$ and $2_1^+$ states are dominated
by $p$-shell components. For the $0_2^+$ state, the $^4$He core is
broken and particles occupy the $sd$ and $pf$ shells.  Similarly, the
$2_2^+$ and $1_1^+$ states exhibit strong core polarization and $sd$
components. The effective interaction will reproduce best those states
that have largest overlap with the model space. Thus, the effective
interaction works well for the low-lying states whose wave function
are dominated by $p$-shell components. Our results are shown in
Fig.~\ref{fig:he6}. The EOM-CC result is from
Ref.~\cite{jansen2011}. In that approach, $^6$He is computed as a
two-particle attached system of $^4$He, including $2p$-$0h$ and
$3p$-$1h$ configurations. The resulting $^6$He ground-state energy is
underbound by about 0.3~MeV compared to FCI.  The SMCC(2b) $0^+$
ground state is about 150~keV over bound compared with FCI, and the
gap between the $0^+$ and $2^+$ states is about 0.6~MeV too large. The
correction of $[[\chi,S]_{3b},S_{2b}]_{2b}$ that enters SMCC(3b-diag)
is very small, and this is in contrast to the closed-shell calculation
of $^4$He. The SMCC(3b-od) calculation including both
$[[\chi,S]_{3b},S]_{2b}$ and perturbative $S_{3b}$ almost reproduces
the EOM-CC results. This demonstrates that in open-shell calculations
the elimination of off-diagonal induced three-body terms is crucial
both for the binding energy and the energy spectrum. The VS-IMSRG(2)
results here are similarly overbound as SMCC(2b).
   
\begin{figure}[ht]
\includegraphics[width=0.5\textwidth]{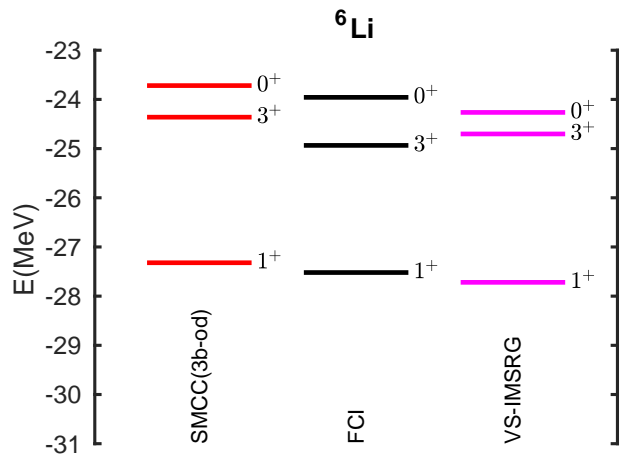}
\caption{(Color online) Spectrum of $^6$Li computed with various
  methods (FCI and VS-IMSRG) and compared to the SMCC(3b) result. SMCC(3b)
  include both three-body interaction diagonal and off-diagonal part.}
\label{fig:li6}
\end{figure}

Fig.~\ref{fig:li6} compares the SMCC(3b-od) result of $^6$Li with FCI and
VS-IMSRG. Here, both valence-space methods agree well with FCI.
    
\begin{figure}[ht]
\includegraphics[width=0.5\textwidth]{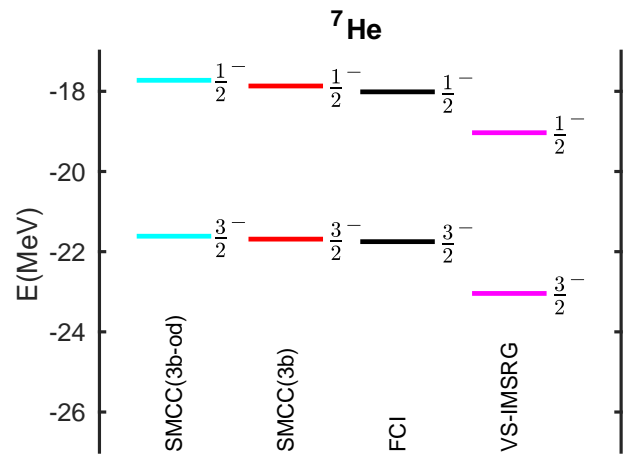}
\caption{(Color online) Spectrum of $^7$He nucleus, computed with various methods
  (FCI, EOM-CC and VS-IMSRG) and compared to the SMCC results. }
\label{fig:he7}
\end{figure}

The calculation of $A=7$ nuclei is interesting as these have three
valence nucleons. Fig.~\ref{fig:he7} shows the SMCC results for
$^7$He and compares them to FCI and VS-IMSRG. The SMCC(3b-od)
calculations both include $[[\chi,S]_{3b},S]_{2b}$ and perturbative
$S_{3b}$, while SMCC(3b) also employs valence three-body forces. Since the valence space
now contains three nucleons, we can see whether the valence three-body
force is important.  For $A=7$, nuclei, it yields a small amount of
additional binding and causes the result to move toward better
agreement with FCI. The VS-IMSRG again agrees roughly with the
SMCC(2b) (not shown), and both poorly reproduce FCI.

\begin{figure}[ht]
\includegraphics[width=0.5\textwidth]{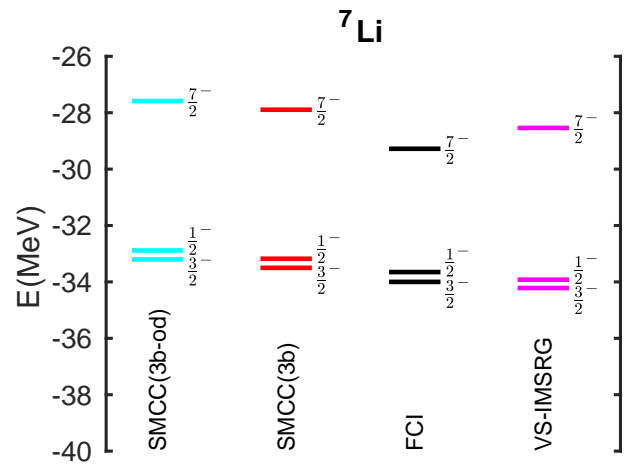}
\caption{(Color online) Same as Fig.~\ref{fig:he7} but for $^7$Li.}
\label{fig:li7}
\end{figure}

For $^7$Li the situation is similar to $^7$He, see
Fig. \ref{fig:li7}), but the SMCC(3b-od) and SMCC(3b) are both
under bound compared to FCI. We anticipate that an
iterative treatment of the three-body forces in valence space would
improve this result.

\begin{figure}[ht]
\includegraphics[width=0.5\textwidth]{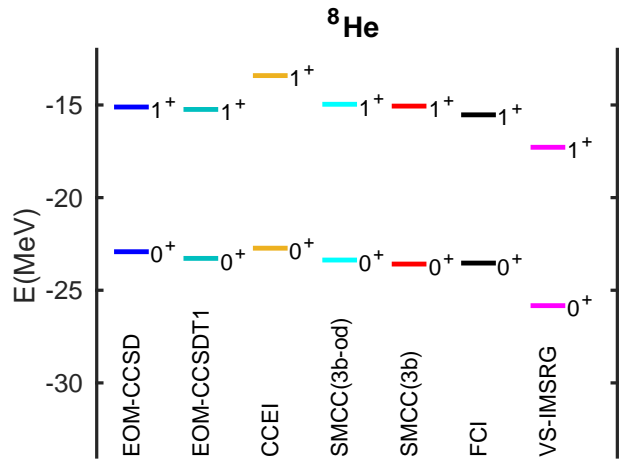}
\caption{(Color online) Spectrum of $^8$He computed with various
  methods (FCI, EOM-CC, CCEI, and IMSRG) and compared to SMCC. The
  SMCC(3b-od) computation employs induced three-body forces that contribute
  to two-body forces, while SMCC(3b) also employs explicit three-body
  forces in the shell-model calculation.}
\label{fig:he8}
 \end{figure}

Finally, we turn to $A=8$ nuclei. The closed sub-shell structure of
$^8$He makes it possible to compute this nucleus within
single-reference CC method. Likewise, $^8$Li can be calculated with
charge-exchange EOM-CC~\cite{ekstrom2014} as a generalized excitation
of $^8$He. Fig.~\ref{fig:he8} shows the comparison of the various
methods in the calculation of $^8$He.  The EOM-CC calculation is close
to the FCI, while results from the recent CC effective interaction
(CCEI) method~\cite{jansen2014,jansen2016} underbinds $^8$He. The
valence three-body force arising from SMCC(3b) makes it more bound on
the order of 200 keV, and results in a good agreement with FCI. The
VS-IMSRG results significantly overbinds, presumably because it also
omits induced three-body effects.

\begin{figure}[ht]
\includegraphics[width=0.5\textwidth]{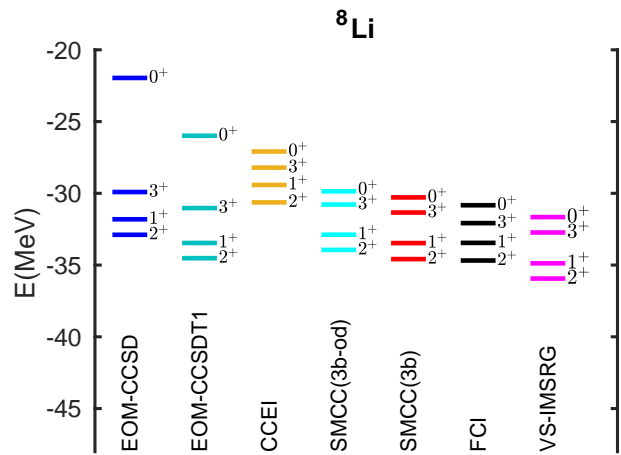}
\caption{(Color online) Same as Fig.~\ref{fig:he8} but for $^8$Li.}
\label{fig:li8}
\end{figure}

Finally, we turn to $^8$Li and show results in Fig.~\ref{fig:li8}. The
SMCC(3b) calculation is significantly improved by including the
valence-space three-body force, on the order of 600 keV, and the low-lying
states are in good agreement with FCI. The VS-IMSRG results reproduce
spectra well, but overbind by about 2 MeV compared to FCI. The
charge-exchange calculation with EOM-CCSDT-1 reproduces the first
three low-lying states well, while the higher-lying $0^+$ probably
lacks correlation energy from neglected higher order particle-hole
excitations.  Again, our results show that SMCC(3b) is an accurate tool for
energy levels.

\begin{figure}[ht]
\includegraphics[width=0.5\textwidth]{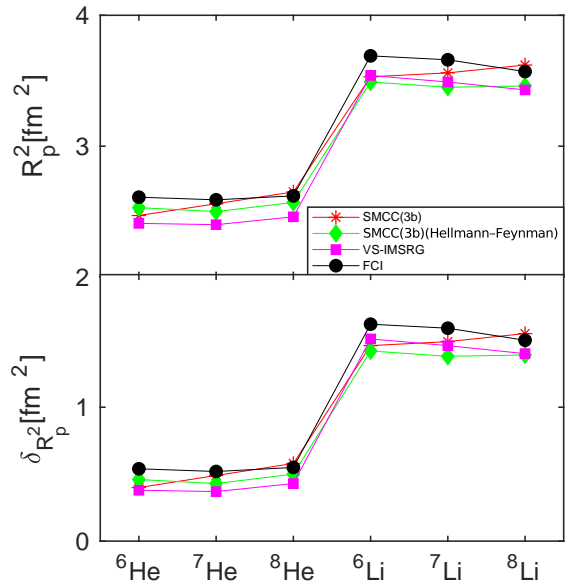}
\caption{(Color online) Squared point proton radii and isotope
  shift($\delta_{R^2_p}=R^2_p-R^2_p(^4\text{He})$) of selected
  $p$-shell nuclei calculated using the SMCC(3b), the Hellmann-Feynman
  theorem [based on SMCC(3b)], the VS-IMSRG, and FCI.}
\label{fig:radii}
\end{figure}

We also calculated the charge radii of He and Li isotopes with the
approximate treatment of the left wave function. The core component is
calculated in the CCSD approximation, which contains the zero-body part of
the operator and the correction from the left (so called $\Lambda$)
wavefunction. Fig.~\ref{fig:radii} shows the calculated charge radii
from SMCC(3b) and VS-IMSRG and compares them to FCI results. We also
calculated the charge radii using the Hellmann-Feynman method on top
of the SMCC, though this approach is not correct because of the
bi-variational structure of the CC energy functional. We note that the
SMCC(3b) approach is closer to the FCI results than VS-IMSRG(2), but
both methods are not as accurate as one would wish.

%% file: summary.tex
\section{Summary}
We proposed a new non-perturbative microscopic method to derive
effective interactions for the shell-model within the framework of
coupled-cluster theory. The resulting shell-model coupled-cluster
(SMCC) method is a promising route to perform precise calculations of
open-shell nuclei with a large number of valence nucleons. Using
renormalization group techniques, induced many-body forces generated
by the similarity transformation are included up to the three-body
level. The effective interaction is applied to selected $p$-shell
nuclei, and we obtained a good agreement with FCI results for energies
and point-proton radii. We demonstrated that the inclusion of induced
three-body forces is important for a precise computation of binding
energies and spectra. The proposed method is a straightforward
extension of the single reference coupled-cluster theory to the
multi-reference system. It keeps the size-extensivity of the single
reference coupled-cluster method, can be systematically improved, and
is capable of dealing with more complex medium-mass nuclei.